\documentclass[journal,comsoc]{IEEEtran}
\usepackage[utf8]{inputenc}
\usepackage[T1]{fontenc}

\usepackage[font=scriptsize]{caption}
\usepackage{multirow}
\usepackage[pdftex]{graphicx}
\usepackage{amsmath}
\usepackage[nocomma]{optidef}
\usepackage[colorinlistoftodos]{todonotes}
\usepackage[cmintegrals]{newtxmath}
\usepackage{xcolor}

\usepackage{subcaption}
\usepackage[english]{babel}

\usepackage{algorithm} 
\usepackage{algpseudocode}
\usepackage{float}

\usepackage{amsfonts}
\usepackage{lipsum}

\title{\vspace{0.0cm}eMBB-URLLC Resource Slicing: A Risk-Sensitive Approach}

\author{Madyan Alsenwi\thanks{Madyan Alsenwi, A. K. Bairagi, and C. S. Hong are with the Department of Computer Science and Engineering, Kyung Hee University, Yongin 446-701, South Korea (email: \{malsenwi, anupam, cshong\}@khu.ac.kr).}, Nguyen H. Tran\thanks{ N. H. Tran is with School of Computer Science, The University of Sydney, NSW 2006, Australia (e-mail: nguyen.tran@sydney.edu.au).}, \IEEEmembership{Senior Member, IEEE}, Mehdi Bennis\thanks{M. Bennis is with the Department of Communications Engineering, University of Oulu, FI-90014 Oulu, Finland, and also with the Department of Computer Engineering, Kyung Hee University, Seoul, South Korea (e-mail: bennis@ee.oulu.fi)}, \IEEEmembership{Senior Member, IEEE}, Anupam Kumar Bairagi, \IEEEmembership{Member, IEEE}, and Choong Seon Hong, \IEEEmembership{Senior Member, IEEE} \vspace{-1cm}

}
\begin{document}

%
%
\maketitle
\begin{abstract}
	Ultra Reliable Low Latency Communication (URLLC) is a 5G New Radio (NR) application that requires strict reliability and latency. URLLC traffic is usually scheduled on top of the ongoing enhanced Mobile Broadband (eMBB) transmissions (\textit{i.e.,} puncturing the current eMBB transmission) and cannot be queued due to its hard latency requirements. In this letter, we propose a risk-sensitive based formulation to allocate resources to the incoming URLLC traffic while minimizing the risk of the eMBB transmission (\textit{i.e.,} protecting the eMBB users with low data rate) and ensuring URLLC reliability. Specifically, the Conditional Value at Risk (CVaR) is introduced as a risk measure for eMBB transmission. Moreover, the reliability constraint of URLLC is formulated as a chance constraint and relaxed based on Markov's inequality. We decompose the formulated problem into two subproblems in order to transform it into a convex form and then alternatively solve them until convergence. Simulation results show that the proposed approach allocates resources to the incoming URLLC traffic efficiently while satisfying the reliability of both eMBB and URLLC.   
\end{abstract}
\IEEEpeerreviewmaketitle
\vspace{-0.3cm}
\section{Introduction}
\IEEEPARstart{T}{he} upcoming Fifth Generation (5G) New Radio (NR) is designed to support three major types of traffic: enhanced Mobile Broad Band (eMBB), massive Machine Type Communications (mMTC), and Ultra Reliable Low Latency Communications (URLLC). While eMBB is an extension of the LTE-Advanced service whose objective is to maximize the peak data rate, mMTC is designed to support a large number of Internet of Things (IoT) devices sending small data sporadically during the active phase only. On the other hand, URLLC is designed to support services that require high level of reliability and low latency. According to the Third Generation Partnership Project (3GPP), the main objective of URLLC is to minimize the latency down to $1 ms$ while ensuring packet error rates of less than $10^{-5}$ \cite{3GPP2018}. These requirements are critical for applications such as industrial IoT,  autonomous vehicles, and virtual reality \cite{bennis2018ultra,Popovski2018}.


The arrival URLLC traffic is immediately transmitted due to its hard latency requirements which may overlap onto previously allocated eMBB transmissions. Allocating resources to the critical URLLC traffic based on a formulation that aims to maximize the total average data rate of eMBB users hinders eMBB users with low data rate and protects higher data rate users. Therefore, we depart from the classical average-based formulation and instead capture the tail of the rate distribution. Considering this risk for eMBB transmission protects low data rate eMBB users and ensures reliable eMBB transmission.

Recently, studies focusing on URLLC have gained attention in both academia and industry. The authors in \cite{Popovski2018} discuss the principles for achieving URLLC and describe several building blocks of framing, use of diversity, and access topology. Authors in \cite{Anand2017b} consider a linear model, convex model, and threshold model for eMBB data rate loss associated with URLLC traffic. Authors in \cite{Pedersen2018d} propose a punctured scheduling approach for transmission of low latency communication (LLC) traffic multiplexed on a shared channel with eMBB. In this work, downlink resource slicing for URLLC and eMBB traffics based on puncturing is considered. In summary, our contributions are as follows: \begin{itemize}
	\item We first formulate a risk-sensitive optimization problem to find the probability of each eMBB user being punctured by URLLC downlink traffic. In contrast to the classical average-based formulation, the risk-sensitive formulation captures the tail of eMBB rate distribution ensuring reliability by protecting users with low data rate. We use the Conditional Value at Risk (CVaR) as a risk measure. Furthermore, we formulate the reliability constraint of URLLC as a chance constraint. In addition, RBs are allocated to eMBB users based on formulation that guarantees proportional fairness.
	\item Due to the non-convexity of the formulated problem, we decompose it into two subproblems: \textit{1) eMBB users scheduling and 2) URLLC placement problem}. The eMBB users scheduling problem is an integer programming problem. Therefore, we relax it to a convex optimization  problem whose solution is within a constant approximation from the optimal. Furthermore, the Markov's inequality is leveraged to relax the URLLC chance constraint of the URLLC placement problem into a linear form. Therefore, the resulting URLLC placement problem is a transformed convex optimization problem for a given eMBB users scheduling which can be solved efficiently by the Base Station (BS). The two problems are then alternatively maximized until convergence.    
	\item Simulation results show that the proposed approach allocates resources to the incoming URLLC traffic while ensuring the reliability of both eMBB and URLLC. The results show that our proposed approach keeps the eMBB reliability higher than $90\%$ for different URLLC traffics. Moreover, the results show the tradeoff between eMBB data rate and URLLC reliability.  
\end{itemize} 
\vspace{-0.4cm}
\section{System Model and Problem Formulation}
\subsection{System Model}
\begin{figure}[t!]
	\centering
	\includegraphics[width=9cm,height=6cm]{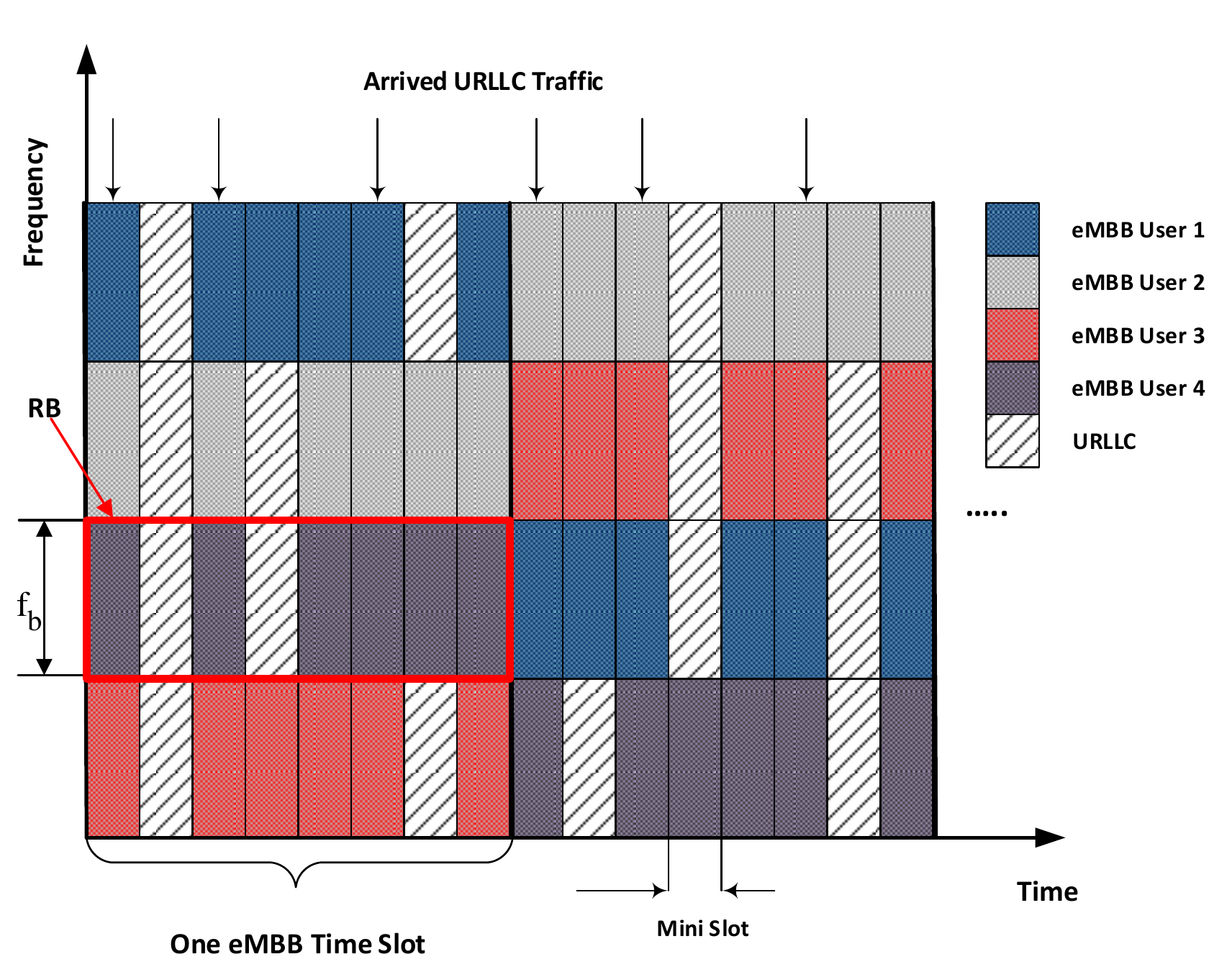}
	\captionof{figure}{Multiplexing of URLLC/eMBB traffics in 5G NR}
	\label{system model}\vspace{-0.2cm}
\end{figure}

We consider the downlink transmissions of a BS with a set of eMBB users denoted by $\mathcal{U}=\{1, 2, ..., U\}$. The BS transmits on a system bandwidth, which is divided into a set of RBs\footnote{5G NR permits a large number of block shapes varying from 15 kHz to 480 kHz} denoted by $\mathcal{B}=\{1, 2, ..., B\}$. The time domain is divided into equally spaced time slots with one millisecond time duration. Each time slot is further divided into minislots\footnote{In 3GPP, the formal term for a ‘slot’ is eMBB Transmit Time Interval (TTI), and a ‘minislot’ is a URLLC TTI.} denoted by $M$ in order to achieve the latency requirement of URLLC \cite{3GPP2018}. RBs are allocated to eMBB users at the slot boundary. However, the sporadic URLLC traffic may arrive during the time slot whose RBs are already allocated to different eMBB users and cannot be delayed to the next time slot due to its hard latency requirements. Therefore, the arriving URLLC traffic is scheduled immediately to transmit in the next minislot as shown in Fig. \ref{system model}. The BS allocates zero transmission power for the overlapped eMBB, which is referred to as \textit{Puncturing} \cite{3GPP2018, Anand2017b}. In this letter, we allocate RBs to eMBB users based on a proportional fair formulation at the boundary of each time slot. Then, we calculate the URLLC placement strategy (\textit{i.e.,} which RBs to \textit{puncture/superpose}) that minimizes the impact on eMBB transmissions.

We consider that $\varphi_u=\sum_{b\in\mathcal{B}}f_bI_{b,u}$ determines the bandwidth allocated to eMBB user $u$, where $f_b$ is the bandwidth of the $b^{th}$ RB and $I_{b,u}$ is the eMBB user scheduling indicator vector. Let the random variable $L_m$ denote the incoming URLLC traffic at minislot $m$ of the considered time slot. Here, $L_m$ can be represented as a Bernoulli distribution with success probability $p$ (\textit{i.e.,} the probability of URLLC transmission at a minislot). 

The loss of eMBB user data rate at each time slot is directly proportional to the punctured resources of that user \cite{Anand2017b}. Let $\omega_u$ define the weight of puncturing the eMBB user $u$, \textit{i.e.,} $\omega_{u}\in[0,1] \;\forall u\in\mathcal{U}$, and $\boldsymbol{\omega}=(\omega_1, \omega_2, ..., \omega_U)$. The users with higher $\omega_{u}$ have higher probability to be punctured by URLLC traffic. Therefore, the instantaneous data rate of an eMBB user $u$ with puncturing weight $\omega_u$ and frequency resources $\varphi_u$ can be approximated based on the Shannon capacity model as follows:\vspace{-0.3cm}
\begin{equation}
R_{u}=\varphi_{u}\bigg(1-\omega_u\times\frac{L}{L_{max}}\bigg)\log_2\Big(1+\frac{P_ug_u}{N_0}\Big), 
\label{impacted data rate of one eMBB user}
\end{equation}
where $L=\sum_{m=1}^{M}L_m$ follows the binomial distribution with parameters $M$ and $p$, $L_{max}$ is the maximum URLLC traffic that can be served at a time slot (\textit{i.e.,} $L_{max}$ equals to the system capacity), $P_u$ and $g_u$ are the transmission power and channel gain of the user $u$ respectively, and $N_0$ is the noise power. The term $(\omega_u\times L/ L_{max})$ is an approximation of the punctured resources of eMBB user $u$ by URLLC traffic, where $L/ L_{max}$ is the normalized URLLC traffic and $\varphi_{u}$ represents the total resources of eMBB user $u$. URLLC traffic exceeding system capacity are blocked thus $L\leq L_{max}$ holds almost surely.
\vspace{-0.3cm}
\subsection{Problem Formulation}
The URLLC scheduler aims to find the URLLC placement weight vector $\boldsymbol{\omega}$ to schedule the arriving URLLC traffic while considering the reliability of eMBB transmissions. Calculating the URLLC placement weight vector based on a formulation that aims to maximize the total average data rate of eMBB users impacts the eMBB users with low data rate while protecting the users with high data rate. Therefore, the proposed URLLC scheduler considers the risk on eMBB transmissions to protect eMBB users with low data rate. The obtained URLLC placement weight $\boldsymbol{\omega^{*}}$ distributes the URLLC traffic among eMBB users such that users with bad channel conditions are considered. In this work, the CVaR, which is also known as the mean excess loss, is used as a risk measure since it captures the tail of the eMBB users data rate. The CVaR provides the average of potential loss that exceed the Value-at-Risk (VaR). The $\alpha-\text{VaR}$ is the $\alpha-$percentile of distribution of a random variable (\textit{i.e.,} a smallest value such that the probability that a random variable is smaller or equals to this value is greater than or equal to $\alpha$) and given by \cite{Rockafellar2000}\vspace{-0.1cm}:
\begin{equation}
\text{VaR}_{\alpha}(R)=\arg \inf\limits_{\gamma}\{\gamma:P(R>\gamma)\leq\alpha\},
\label{VaR}
\end{equation}
where $R=\sum_{u\in \mathcal{U}}R_u$, and $\alpha \in (0, 1)$. The CVaR function is defined as the expectation of the $\alpha$ fraction of the worst outcomes of $R$: 
\begin{equation}
\text{CVaR}_{\alpha}(R)=\mathbb{E}[R|R>\text{VaR}_{\alpha}(R)].
\label{CVaR Definition}
\end{equation}
Moreover, we have that \cite{Rockafellar2000}:
\begin{align}
\phi_{\alpha}(R, \gamma):=\gamma+\frac{1}{1-\alpha}\mathbb{E}\big[(R-\gamma)^+\big],
\label{phi funciton}
\end{align}
where $(x)^+=\max(0, x)$,
and the $\text{CVaR}_{\alpha}$ of the random variable $R$ can be determined as
\begin{equation}
\text{CVaR}_{\alpha}(R)=\min\limits_{\gamma\in \mathbb{R}}\phi_{\alpha}(R, \gamma).
\label{CVaR}
\end{equation}
As $\mathbb{E}[L] = Mp$, we can rewrite the $\mathbb{E}[R]$ in \eqref{phi funciton} as follows:
\begin{align}
\mathbb{E}[R]=\sum\limits_{u\in \mathcal{U}}\varphi_{u}\bigg(1-\omega_u\times\frac{Mp}{L_{max}}\bigg)\log_2\Big(1+\frac{P_ug_u}{N_0}\Big).   
\end{align}

Let $\mathcal{C}=\{1, 2, ..., C\}$ be the set of all URLLC users, $R_c$ is the data rate of a URLLC user $c$,  and $R_{urllc}=\sum_{c\in \mathcal{C}}R_c$ represents the data rate of all URLLC users. Therefore, the outage probability of URLLC is given as: 
\begin{equation}
P(E)=Pr\big[R_{urllc} \leq L\big],
\label{outage probability}
\end{equation}
where $R_{urllc}$ is given by:
\begin{equation}\vspace{-0.1cm}
R_{urllc}(\theta)=\sum\limits_{c\in\mathcal{C}}\sum\limits_{u\in \mathcal{U}}\frac{\theta_u}{C}\log_2\Big(1+\frac{P_{c}g_{c}}{N_0}\Big).
\label{urllc_rate_per_user}
\end{equation}\vspace{-0.1cm}

In \eqref{urllc_rate_per_user}, $\theta_u=(\phi_{u}\times \omega_u \frac{L}{L_{max}})$ represents the punctured resources from the eMBB user $u$ to URLLC traffic, $P_{c}$ and $g_c$ are the transmission power and channel gain of URLLC user $c$, respectively. Here, we consider that the total punctured resources to URLLC traffic is divided equally between the URLLC users.
 
RBs are allocated to eMBB users at the beginning of each time slot by solving an optimization problem that guarantees proportional fairness. The proportional fair resource allocation is modeled as maximizing the sum of logarithms of data rates \cite{Girici2010}. Therefore, the final optimization problem of both eMBB and URLLC schedulers can be formulated as follows:
\begin{maxi!}[2]                 
	{\boldsymbol{\omega}, \boldsymbol{I}}                               
	{\sum\limits_{u\in \mathcal{U}}\log\big(\mathbb{E}[R_u]\big)-\beta\Big(\text{CVaR}_{\alpha}\big(R\big)\Big) \label{objective function} }  
	{\label{eMBB and URLLC optimization problem}}             
	{}                                
	\addConstraint{ Pr\big[R_{urllc}\leq L\big]}{\leq \epsilon,  \label{urllc_const_main}}
	\addConstraint{\sum\limits_{u\in\mathcal{U}}I_{b,u}}{\leq 1,\;\; \forall \; b\in \mathcal{B}, \label{const2}} 
	\addConstraint{I_{b,u}}{\in \{0, 1\}, \;\; \forall u\in \mathcal{U}\; and\; b \in \mathcal{B},}
	\addConstraint{0\leq\omega_{u}}{\leq 1, \;\; \forall u \in \mathcal{U},}
\end{maxi!}\vspace{-0.4cm} \\
where $\beta\in[0,1]$ is the weight of the CVaR function, and $\epsilon$ denotes the reliability of URLLC and takes a small value. The above optimization problem seeks both the optimum resource allocation matrix $\boldsymbol{I^*}$ for eMBB users and the optimum URLLC placement weight vector $\boldsymbol{\omega^*}$. The objective function (\ref{objective function}) is formulated based on a sum-log formulation to ensure proportional fairness when allocating the resources to eMBB users. Furthermore, the CVaR function minimizes the risk of eMBB users when allocating resources to URLLC traffic.
\vspace{-0.1cm}
\section{Proposed Solution}
The original problem (\ref{eMBB and URLLC optimization problem}) is a mixed-integer nonlinear programming (MINLP). To find a global optimum solution, we need to search the space of feasible URLLC placement weights with all possible combinations of eMBB user scheduling. This may require exponential-complexity to solve. To solve this problem efficiently, we decompose the original problem into two subproblems: \textit{1) eMBB users scheduling}, and \textit{2) URLLC placement problem.}
\vspace{-0.3cm}
\subsection{Resource Allocation for eMBB users}
For any fixed feasible URLLC weight vector $\boldsymbol{\omega}$, the original problem (\ref{eMBB and URLLC optimization problem}) can be presented as follows: 
\begin{maxi!}[2]                 
	{\boldsymbol{\boldsymbol{I}}}                               
	{\sum\limits_{u\in \mathcal{U}}\log\big(\mathbb{E}[R_u]\big)-\beta\Big(\text{CVaR}_{\alpha}\big(R\big)\Big)\label{objective function of eMBB scheduler} }  
	{\label{eMBB Scheduler optimization problem}}             
	{}                                
	\addConstraint{\sum\limits_{u\in\mathcal{U}}I_{b,u}}{\leq 1,\;\; \forall \; b\in \mathcal{B}, }
	\addConstraint{I_{b,u}}{\in \{0, 1\}, \;\; \forall u\in \mathcal{U}\; and\; b \in \mathcal{B}.}
\end{maxi!}\vspace{-0.1cm}
\begin{algorithm}[t!]
	\caption{Resource Allocation Strategy for eMBB and URLLC Traffic}\label{Algorithm1}
	\begin{algorithmic}[1]
		\State \text{URLLC weight vector initialization}
		\State \textbf{repeat:}
		\State \hspace{1.1cm}\text{eMBB users scheduling (\ref{eMBB Scheduler optimization problem_relaxed})}
		\State \hspace{1.1cm}\text{URLLC placement problem (\ref{URLLC optimization problem2})}
		\State \textbf{until }$\boldsymbol{I}$ $\&$ $\boldsymbol{\omega}$  \text{converge or max \# of iterations is reached}
	\end{algorithmic}
\label{algorithm1}
\end{algorithm}
The integer programming problem (\ref{eMBB Scheduler optimization problem}) can be relaxed to a convex optimization  problem whose solution is within a constant approximation from the optimal. Then, the fractional solution is rounded to get a solution to the original integer problem. The randomization of $R_u$ comes from the URLLC part only (\textit{i.e.,} we consider that the BS can estimate the channel gain of all users). Therefore, considering only the RBs allocation to eMBB users problem leads to a deterministic variable of $R_u$. In this case, the CVaR part can be removed from the objective function. Accordingly, the optimization problem (\ref{eMBB Scheduler optimization problem}) can be approximated as follows: 
\begin{maxi!}[2]                 
	{\boldsymbol{\boldsymbol{I}}}                               
	{\sum\limits_{u\in \mathcal{U}}\log\Big(\mathbb{E}\big[R_u\big(\phi(I)\big)\big]\Big)\label{objective function of eMBB scheduler_relaxed} }  
	{\label{eMBB Scheduler optimization problem_relaxed}}             
	{}                                
	\addConstraint{\sum\limits_{u\in\mathcal{U}}I_{b,u}}{\leq 1,\;\; \forall \; b\in \mathcal{B}, }
	\addConstraint{0\leq I_{b,u}}{\leq 1, \;\; \forall u\in \mathcal{U}\; and\; b \in \mathcal{B}.}
\end{maxi!}
The convex optimization problem (\ref{eMBB Scheduler optimization problem_relaxed}) can be solved by applying the Karush-Kuhn-Tucker (KKT) conditions.
\vspace{-0.4cm}
\subsection{URLLC Scheduler for a Given eMBB User Scheduling Matrix}\vspace{-0.1cm}
For any given eMBB user scheduling $\boldsymbol{I}$, the original problem (\ref{eMBB and URLLC optimization problem}) can be reduced to the following URLLC placement strategy problem:  \vspace{-0.3cm}
\begin{maxi!}[2]                 
	{\boldsymbol{\omega}, \boldsymbol{\gamma}}                               
	{\sum\limits_{u\in \mathcal{U}}\log\Big(\mathbb{E}\big[R_u(\omega)\big]\Big)-\beta\bigg(\gamma+\frac{1}{(1-\alpha)}(\mathbb{E}[R]-\gamma)^+\bigg)  \label{objective function of URLLC placement1} }  
	{\label{URLLC optimization problem1}}             
	{}  
	\addConstraint{ Pr\big[R_{urllc}(\omega)\leq L\big]}{\leq \epsilon, \;\; 0\leq\omega_{u}\leq 1,\;\; \forall u \in \mathcal{U}.\label{urllc_const_2}}                              
\end{maxi!}\vspace{-0.0cm}
We use the Markov's Inequality to represent the chance constraint (\ref{urllc_const_2}) as a linear constraint:
\begin{equation}\vspace{-0.0cm}
Pr\big[R_{urllc}\leq L\big]
\leq \frac{\mathbb{E}[L]}{R_{urllc}}.
\end{equation}
\begin{figure*}[t!]
	\centering
	\begin{subfigure}[b]{0.3\textwidth}
		\includegraphics[width=\textwidth, height=1.5in]{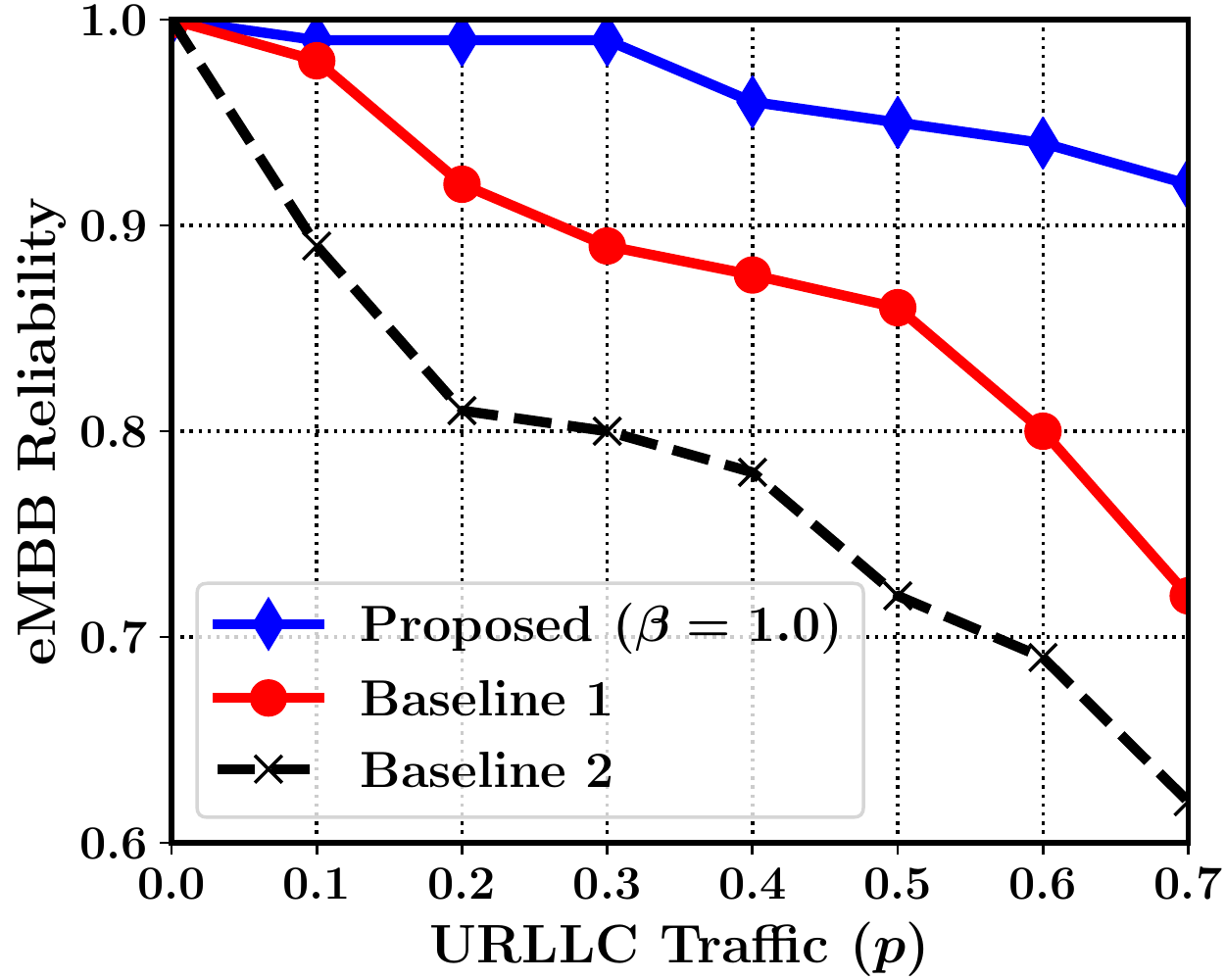}
		\caption{\footnotesize eMBB reliability with URLLC traffic}
		\label{eMBB_rliability}
	\end{subfigure}
	\begin{subfigure}[b]{0.3\textwidth}
		\includegraphics[width=\textwidth, height=1.5in]{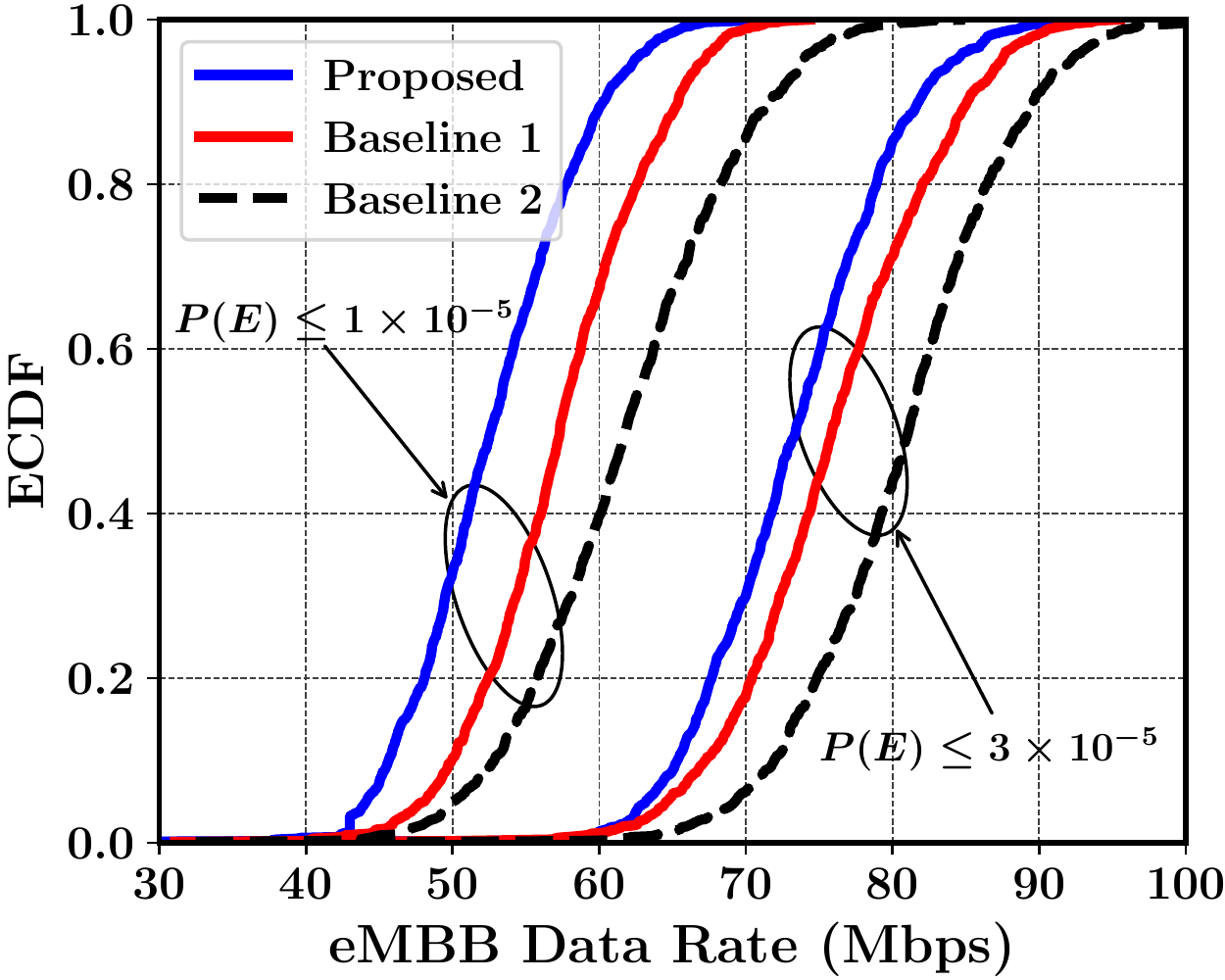}
		\caption{\footnotesize eMBB rate for different URLLC reliability}
		\label{eMBB_rate_epsilon}
	\end{subfigure}
	\begin{subfigure}[b]{0.3\textwidth}
		\includegraphics[width=\textwidth, height=1.5in]{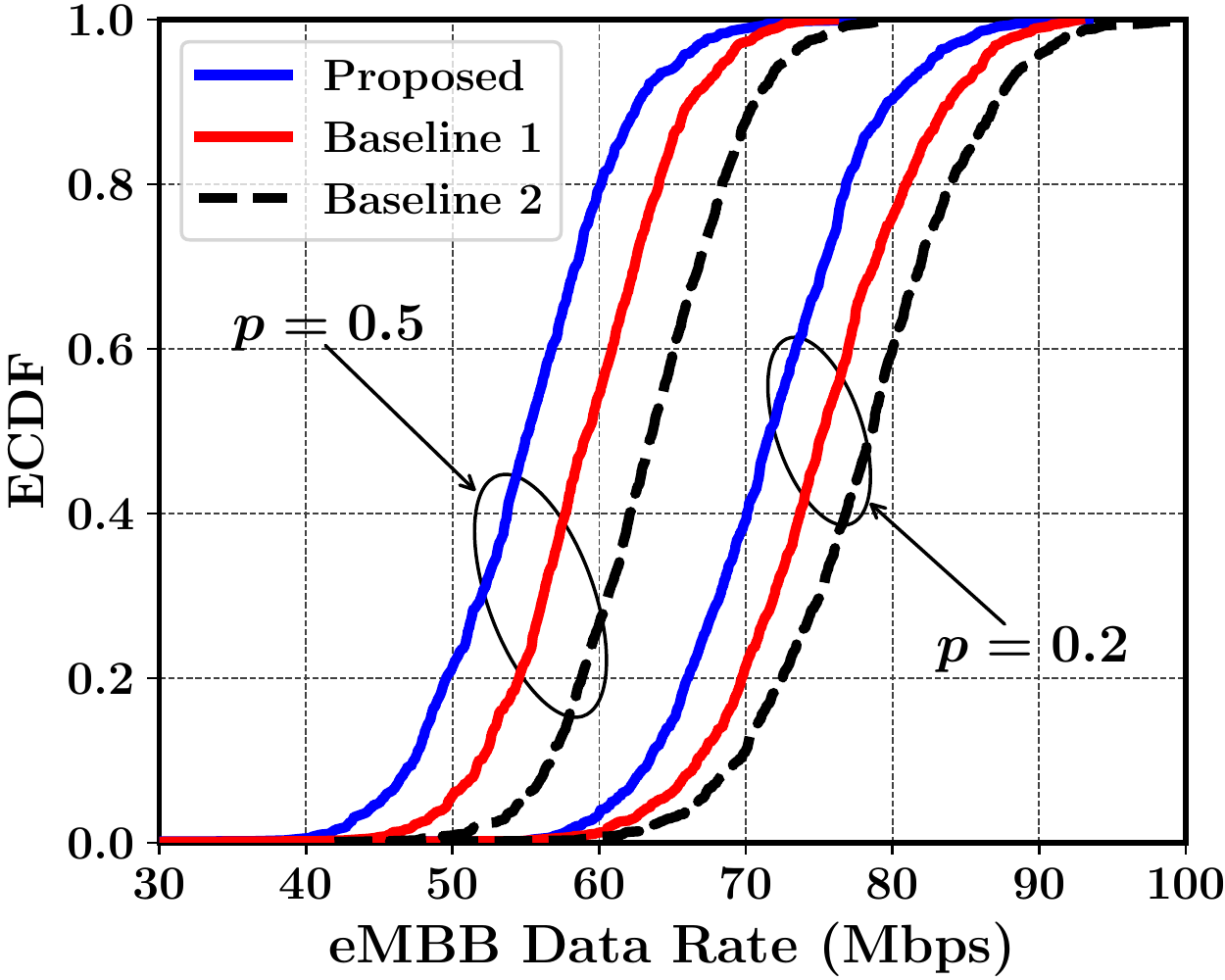}
		\caption{\footnotesize eMBB data rate for different URLLC traffic}
		\label{eMBB_rate_p}
	\end{subfigure}
	~ 
	~ 
	\vspace{0.3cm}
	\caption{URLLC/eMBB tradeoffs.}
	\label{ECDF_eMBB}
	\vspace{-0.3cm}
\end{figure*}   
Furthermore, we introduce variable $z$ to replace $(\mathbb{E}[R]-\gamma)^+$. This is achieved by imposing the constraints $z\geq \mathbb{E}[R]-\gamma$ and $z\geq 0$:\vspace{-0.2cm}
\begin{maxi!}[2]                 
	{\boldsymbol{\omega}, \boldsymbol{z}, \boldsymbol{\gamma}}                               
	{\sum\limits_{u\in \mathcal{U}}\log\Big(\mathbb{E}\big[R_u(\omega_u)\big]\Big)-\beta\bigg(\gamma +\frac{z}{(1-\alpha)}\bigg) \label{objective function of URLLC placement2} }  
	{\label{URLLC optimization problem2}}             
	{}      
	\addConstraint{\mathbb{E}[R_{urllc}(\boldsymbol{\omega})]} {\geq \frac{MP}{\epsilon}, \;\; z\geq \mathbb{E}[R(\boldsymbol{\omega})]-\gamma, \label{relaxed constraint}}   
	\addConstraint{z\geq 0, \;\;}{0\leq\omega_{u}\leq 1, \;\; \forall u \in \mathcal{U}.}
\end{maxi!}\vspace{0.0cm}
For any given eMBB user scheduling, the above problem is a convex optimization problem which can be solved efficiently by the BS. The BS only needs to estimate the wireless channel gains of eMBB users that are time varying in each time slot and the other parameters remain constant for sufficiently long period of time. Therefore, the BS solves (\ref{URLLC optimization problem2}) and broadcasts the solution to both eMBB and URLLC users. Alg. \ref{Algorithm1} describes the proposed approach.
\vspace{-0.3cm}
\section{Performance Evaluation}
The performance of the proposed approach is evaluated in terms of eMBB reliability and total sum-rate of eMBB. We evaluate the performance of the proposed risk-sensitive based approach for different values of the parameters $p$, and $\epsilon$ and compare it to : 
\textit{1) Baseline 1 (Proposed with $\beta=0$):} setting $\beta=0$ omits the CVaR part from the objective function and allocates resources to URLLC based on maximizing the sum-log of eMBB users data rate (proportional-fair allocation). \textit{2) Baseline 2 (linear sum rate maximization):} the BS allocates resources to URLLC traffic such that the linear sum of eMBB users data rate (average-based) is maximized.
\begin{figure}[t!]
	\includegraphics[width=3.487in, height=1.3in]{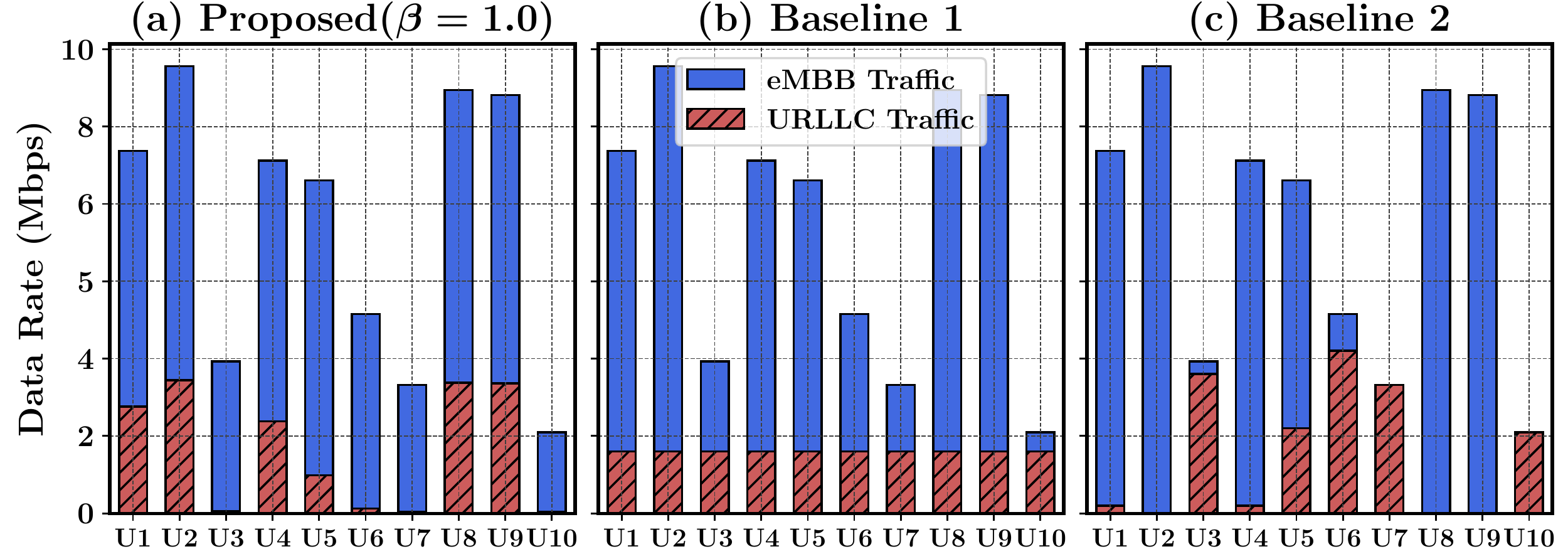}
	\caption{Data rate of eMBB users and distributed URLLC traffic}
	\label{eMBB_URLLC}
	\vspace{-0.2cm}
\end{figure}

Fig. \ref{ECDF_eMBB}(a) shows the impact of URLLC traffic on eMBB reliability, defined as the fraction of eMBB users who achieve the target rate $R_0$. It is shown that for a given target rate $R_0=2 Mbps$ the proposed approach with $\beta=1$ guarantees higher reliability as compared to both \textit{Baseline 1} and \textit{Baseline 2}. Moreover, an increase in URLLC traffic (higher $p$) is shown to decrease the eMBB reliability. Fig. \ref{ECDF_eMBB}(b) and Fig. \ref{ECDF_eMBB}(c) show the Empirical Cumulative Distribution Function (ECDF) of the total eMBB data rate for different values of $\epsilon$ and $p$ calculated over $1000$ time slots. Fig. \ref{ECDF_eMBB} (b) shows the trade off between the URLLC reliability and eMBB data rate. In this figure, increasing $\epsilon$ leads to lower reliability level for URLLC traffic while improving the eMBB data rate. However, decreasing $\epsilon$ means smaller $P(E)$ (\textit{i.e.,} $P(E)\leq \epsilon$) and higher reliability for URLLC, which has more impact on the eMBB data rate. In addition, Fig. \ref{ECDF_eMBB} (b) shows that \textit{Baseline 2} gives higher data rate since its objective is to maximize the total eMBB data rate without considering eMBB reliability, giving eMBB users with low data rate higher probability to be punctured by URLLC traffic. On the other hand, Fig. \ref{ECDF_eMBB} (c) shows the impact of URLLC traffic on the eMBB data rate. As shown in this figure, increasing $p$ means having more URLLC traffic and this leads to more impacting on the eMBB transmission. We also note that \textit{Baseline 2} gives higher eMBB data rate since its objective is to maximize the sum-rate of eMBB transmissions without considering its reliability.

Finally, Fig. \ref{eMBB_URLLC} shows the data rate of eMBB users and the distributed URLLC traffic over eMBB users. As shown in Fig. \ref{eMBB_URLLC} (a), the proposed risk-sensitive based approach gives the eMBB users with high data rates higher probability to be punctured by URLLC traffic while protecting the eMBB users with bad channel conditions. However, setting $\beta=0$ (Baseline 1) distributes the incoming URLLC traffic equally among all eMBB users since it aims at a proportional-fair as shown in Fig. \ref{eMBB_URLLC} (b). Furthermore, \textit{Baseline 2} gives the eMBB users with low data rate higher probability to be punctured by URLLC traffic since its objective is to maximize the sum eMBB data rate as shown in Fig. \ref{eMBB_URLLC} (c).
\vspace{-0.3cm}
\section{Conclusions}
In this work, we studied the problem of dynamic multiplexing of URLLC traffic by puncturing eMBB resources. The resource scheduling problem is formulated as an optimization problem that aims to maximize the total eMBB data rate while considering the risk of eMBB using the CVaR function as a risk measure. The results have shown that the proposed algorithm protects eMBB users with low data rate by allocating more URLLC traffic to eMBB users with higher data rate.
\nocite{*}
\vspace{-0.3cm}
\bibliographystyle{IEEEtran}
\bibliography{References}
	
\end{document}